\documentclass[prl,twocolumn,showpacs]{revtex4}

\usepackage{amsfonts}
\usepackage{amssymb}
\usepackage{graphicx}
\usepackage{pstricks}

\newcommand{\be}{\begin{equation}}
\newcommand{\bea}{\begin{eqnarray}}
\newcommand{\ee}{\end{equation}}
\newcommand{\eea}{\end{eqnarray}}
\newcommand{\bpi}{\begin{picture}}
\newcommand{\bce}{\begin{center}}
\newcommand{\epi}{\end{picture}}
\newcommand{\ece}{\end{center}}

\begin{document}

\title{Novel type of CPT violation for correlated EPR states}
%\date{\today}

\author{J. Bernab\'eu$^a$}
\author{N.E. Mavromatos$^{a,b}$}
\author{J. Papavassiliou$^a$}
\affiliation{$^a$Departamento de F\'\i sica Te\'orica and IFIC, Centro Mixto, 
Universidad de Valencia-CSIC,
E-46100, Burjassot, Valencia, Spain. \\
$^b$King's College London, University of London, Department of Physics, 
Strand WC2R 2LS, London, U.K.}

\begin{abstract}

 We discuss modifications to the 
concept of an ``antiparticle'', induced by a breakdown of the CPT symmetry
at a fundamental level, realized within an 
extended class of quantum gravity models.
The resulting loss of particle-antiparticle 
identity in the neutral-meson system 
induces a breaking of the EPR correlation imposed by Bose statistics.
The latter is parametrized by a complex parameter controlling   
the amount of contamination by the ``wrong'' symmetry state.
The physical 
consequences are studied, and  
novel observables of CPT-violation 
in $\phi$ factories are proposed.

\end{abstract}


\preprint{FTUV-03-1015, IFIC/03-45}

\maketitle

%\vspace{1.cm}

The CPT theorem is one of the most profound results of quantum field 
theory~\cite{cpt}. It is a consequence of Lorentz invariance, 
locality, as well as quantum mechanics (specifically unitary evolution
of a system). 
One implication of CPT invariance is the equality of the masses 
between 
particles and antiparticles. 
In this respect, 
the best experimental tests of the CPT symmetry so far 
have been in the neutral Kaon system, where the equality of 
particle -- antiparticle masses has been confirmed to better than one part in $10^{17}$~\cite{pdg}. 
However, this is not the end of the story, given that CPT violation
may manifest itself in many subtle ways, thus motivating  
further experimental searches in various directions. 

The possibility of a violation of CPT invariance has been considered in a 
number of theoretical contexts that go beyond conventional local 
quantum field theory. In several models of  
quantum gravity (QG), for example,
the axioms
of quantum field theory, as well as conventional quantum mechanical
behaviour, may not be maintained~\cite{hawking} in  
the presence of  special field configurations, such as  
wormholes, microscopic (Planck size) black holes, 
and other topologically non-trivial solitonic objects, such as 
{\it geons}~\cite{geons}. 
Such configurations are collectively referred to as 
{\it space time foam}, a terminology coined by 
J.A. Wheeler~\cite{wheeler}, who first conceived 
the idea that 
the structure of {\it quantum} space-time at Planckian scales ($10^{-35}$ m)
may actually be fuzzy, 
characterised by a ``foamy'' nature.  
Given that such ``objects''   
cannot be accessible to low-energy observers,  
it has been argued that
a mixed state description must be 
employed ({\it QG-induced decoherence})~\cite{hawking,ehns},
``tracing'' over them in the context of an effective field theory.  
In the case of microscopic black holes, for example,
the decoherence arises due to the loss of information across microscopic
event horizons, leading to complications in defining 
proper asymptotic state-vectors and thus 
a Heisenberg  scattering matrix.  
As a corollary of this, it has been argued~\cite{wald}
that, in general, CPT invariance in its strong form  
must be abandoned in quantum gravity.
Since in such models the breakdown 
of the CPT symmetry happens at a fundamental 
level, it would imply 
that a proper CPT operator is {\it ill defined}. This in turn would 
lead to possible deviations from standard quantum mechanical evolution of states~\cite{lopez},
which may not be necessarily associated with the mass difference between particle and antiparticle.   
  
In addition to such effects on the quantum mechanical evolution of a state, however, 
the violation of CPT invariance leads to a modified concept of 
what one calls an {\it antiparticle state}. This is an aspect that has not been 
discussed previously, and, as we shall argue in the present article, leads to novel 
observables that could parametrize the CPT violation. 
Usually the antiparticle 
is defined as the state with quantum numbers such that, upon 
interaction with the corresponding particle it produces 
a state with the quantum numbers of the {\it vacuum} (annihilation). 
If the CPT operator is well defined, such a state is obtained by the 
action of this operator on the corresponding particle state. If, however, the operator
is ill defined, the particle and antiparticle  spaces 
should be thought of as 
{\it independent} subspaces of matter states. 
In this case, the usual assumption for {\it identical states}, 
when supplemented by particle-antiparticle conjugation,  
in the case of 
the electrically neutral mesons $K^0$ and ${\overline K}^0$ 
(or $B^0$ and ${\overline B}^0$), 
which requires their
symmetry under the exchange operator ${\cal P}$ 
as a natural consequence of Bose statistics, {\it is relaxed}. This, 
in turn, modifies the description of (neutral) meson entangled states,
and may bring about significant deviations to their EPR correlations. 
The purpose of this paper is to explore these issues, 
and propose novel CPT-violating observables for the $\phi-$ and $B-$ factories.

In conventional 
formulations of {\it entangled} meson  
states~\cite{dunietz,botella,bernabeu}
one imposes the requirement of {\it Bose statistics} 
for the state $K^0 {\overline K}^0$ (or $B^0 {\overline B}^0$), 
which implies that the physical neutral meson-antimeson state 
must be {\it symmetric} under the combined operation $C{\cal P}$,
with $C$ the charge conjugation and 
${\cal P}$ the operator that permutes the spatial coordinates. 
Specifically, assuming 
{\it conservation} of angular momentum, and 
a proper existence of the {\it antiparticle state} (denoted by a bar),
one observes that, 
for $K^0{\overline K}^0$ states which are $C$-conjugates with 
$C=(-1)^\ell$ (with $\ell$ the angular momentum quantum number), 
the system has to be an eigenstate of 
${\cal P}$ with eigenvalue $(-1)^\ell$. 
Hence, for $\ell =1$, we have that $C=-$, implying ${\cal P}=-$.
As a consequence 
of Bose statistics this ensures that for $\ell = 1$ 
the state of two identical bosons is forbidden~\cite{dunietz}.  
As a result, the initial entangled state 
$K^0{\overline K}^0$ produced in a $\phi $ factory 
can be written as:
{\small 
\begin{equation} 
|i> = \frac{1}{\sqrt{2}}\left(|K^0({\vec k}),{\overline K}^0(-{\vec k})>
- |{\overline K}^0({\vec k}),{K}^0(-{\vec k})>\right)
\label{bbar}
\end{equation}}
This is the starting point 
of all formalisms known to date, either 
in the $K$-system~\cite{dunietz,botella} or
in the $B$-system~\cite{bernabeu}, 
including those~\cite{huet}
where the evolution
of the entangled state is described by non-quantum 
mechanical terms, in the formalism of \cite{ehns}. 
In fact, in all these works it has been claimed  
that the expression in Eq.(\ref{bbar})
is actually independent of any assumption about CP, T or CPT symmetries. 

However, as has been alluded above, 
the assumptions leading to 
Eq.(\ref{bbar}) may not be valid if CPT symmetry is violated.
In such a case 
${\overline K}^0$ cannot be considered 
as identical to ${K}^0$, and thus the requirement of $C {\cal P} = +$, imposed 
by Bose-statistics, is relaxed.
As a result, the initial entangled state (\ref{bbar}) 
can be parametrised in general as:
{\small 
\begin{eqnarray} 
|i> &=& \frac{1}{\sqrt{2}}\left(|K^0({\vec k}),{\overline K}^0(-{\vec k})>
- |{\overline K}^0({\vec k}),{K}^0(-{\vec k})> \right)  \nonumber \\
&+& \frac{\omega}{\sqrt{2}}\left(|K^0({\vec k}),{\overline K}^0(-{\vec k})>
 + |{\overline K}^0({\vec k}),{K}^0(-{\vec k})> \right)  
\label{bbarcptv}
\end{eqnarray}}
where $\omega = |\omega| e^{i\Omega}$ 
is a {\it complex} CPT violating (CPTV) parameter, 
associated with the non-identical particle nature 
of the neutral meson
and antimeson states. This parameter describes a {\it novel} phenomenon, 
not included in previous analyses.

Notice that an equation such as the one given in  
(\ref{bbarcptv}) could also be produced  as a result 
of deviations from the laws of quantum mechanics, 
during the initial decay of the $\phi$ or
$\Upsilon$ states. Thus, Eq.(\ref{bbarcptv}) 
could receive contributions from two different effects,
and can be thought off as simultaneously parametrizing 
both of them.
In the present article we will assume that Eq.(\ref{bbarcptv})
arises solely  due to deviations from the identical-particle
nature of the neutral Kaon and Antikaon states, while 
the Hamiltonian evolution of the entangled state is governed entirely  
by the laws of quantum mechanics. Of course, 
in an actual QG decoherening 
situation one may have to invoke non-quantum-mechanical, open-system  
evolution a\' la \cite{ehns,lopez,huet}; however, this lies beyond 
the scope of the present work, and will be addressed elsewhere.  

We now proceed to study the possible consequences of Eq.(\ref{bbarcptv}). 
To this end, we should first express the initial state in terms of 
CP eigenstates, and also in terms of mass eigenstates, which will be useful
when we discuss decays. In terms of CP eigenstates $K_{\pm} = 
\frac{1}{\sqrt{2}}\left( |K^0> \pm |{\overline K}^0> \right)$, 
the initial entangled state (\ref{bbarcptv}) reads
(for definiteness we concentrate from now on on the $\phi$/Kaons case, although our
formalism is generic and applies equally to $B^0$-mesons, {\it etc}):  
{\small 
\begin{eqnarray} 
|i> &=& \frac{1}{\sqrt{2}}\left(|K_-({\vec k}),K_+(-{\vec k})>
- |K_+({\vec k}),K_-(-{\vec k})> \right)  \nonumber \\
&+& \frac{\omega}{\sqrt{2}}\left(|K_+({\vec k}), K_+(-{\vec k})>
- |K_-({\vec k}),K_-(-{\vec k})> \right)  
\label{bpm}
\end{eqnarray}}
Notice the appearance of $K_+K_+$ or $K_-K_-$ combinations, 
as a result of the CPTV parameter $\omega$, which would not exist if the conventional expression
(\ref{bbar}) had been used instead of (\ref{bbarcptv}). 

Let us express now (\ref{bbarcptv}) in terms of 
physical (mass) eigenstates, defined as
$K_S = \frac{1}{\sqrt{1 + |\epsilon_1^2|}}\left(|K_+> 
+ \epsilon_1 |K_->\right)$, 
$K_L = 
\frac{1}{\sqrt{1 + |\epsilon_2^2|}}\left(|K_-> + \epsilon_2 |K_+>\right)$,
where $\epsilon_1, \epsilon_2$ are complex parameters, and such that, 
if CPT invariance of the Hamiltonian is assumed (within a quantum mechanical
framework), $\epsilon_1=\epsilon_2$, otherwise the quantity 
$\delta \equiv \epsilon_1 - \epsilon_2$ parametrizes the CPT violation
within quantum mechanics. It is convenient to use 
the CP-violating parameters $\delta$ and 
$\epsilon \equiv  |\epsilon|e^{i\phi_\epsilon} =
\frac{\epsilon_1 + \epsilon_2}{2} $ 
to parametrize CPT and 
T violation in a quantum mechanical framework.

\begin{figure}[b]
\includegraphics[width=0.4\textwidth]{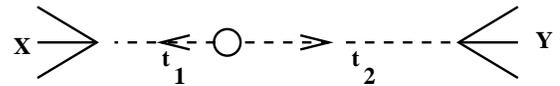}
\caption{A typical amplitude corresponding to the decay 
of, say, a $\phi$ state into final states $X,Y$; $t_i, i=1,2$ denote the 
corresponding time scales for the appearance of the final products of the 
decay.}
\label{amplitude} 
\end{figure}

In terms of such physical eigenstates, the state (\ref{bbarcptv})
is written as (we keep linear terms in the small parameters $\omega$, 
$\delta$, i.e. in the following we ignore higher-order terms 
$\omega \delta$, $\delta^2$  {\it etc.})
{\small 
\begin{eqnarray} 
|i> &=& 
C \bigg[ \left(|K_S({\vec k}),K_L(-{\vec k})>
- |K_L({\vec k}),K_S(-{\vec k})> \right)\nonumber \\  
&+& \omega \left(|K_S({\vec k}), K_S(-{\vec k})>
- |K_L({\vec k}),K_L(-{\vec k})> \right)\bigg]  
\label{bph}
\end{eqnarray}}
with $C = \frac{\sqrt{(1 + |\epsilon_1|^2)
(1 + |\epsilon_2|^2 )}}{\sqrt{2}(1-\epsilon_1\epsilon_2)}
\simeq \frac{1 + |\epsilon^2|}{\sqrt{2}(1 - \epsilon^2)}$.
Notice again the presence of combinations $K_S K_S$ and $K_L K_L$ states,
proportional to the novel CPTV parameter $\omega$.
As we will see, 
such terms become important when one considers decay channels.

Specifically, consider the decay 
amplitude corresponding to the appearance of a final state $X$ 
at time $t_1$ and $Y$ at time $t_2$, as illustrated in  fig. \ref{amplitude}. 
One assumes (\ref{bph}) for the initial two-Kaon system, 
after the $\phi$ decay. The time is set $t=0$ at the moment of the decay.
Denoting the corresponding amplitude by $A(X,Y)$ we have~\cite{dunietz,botella,bernabeu}:  
\be
A(X,Y) = \langle X|K_S\rangle \langle Y|K_S \rangle \, \cal C \,\left( A_1  +  A_2 \right)
\label{axy}
\ee
with
\begin{eqnarray}
 A_1  &=& e^{-i(\lambda_L+\lambda_S)t/2} 
[\eta_X  e^{-i \Delta\lambda \Delta t/2} 
-\eta_Y  e^{i \Delta\lambda \Delta t/2}]\nonumber \\  
A_2  &=&  \omega [ e^{-i \lambda_S t} - 
\eta_X \eta_Y e^{-i \lambda_L t}] 
\end{eqnarray}
the CPT-allowed and CPT-violating parameters respectively, 
and 
$\eta_X = \langle X|K_S\rangle/\langle X|K_L\rangle$ and  
$\eta_Y =\langle Y|K_S\rangle/\langle Y|K_L\rangle$.
Next, one integrates 
the square of the amplitude 
over all accessible times $t= t_1 + t_2$,
keeping the difference $\Delta t = t_2 - t_1$ as constant.   
This defines the ``intensity'' $I(\Delta t)$:
\begin{eqnarray} 
I (\Delta t) \equiv \frac{1}{2} \int_{|\Delta t|}^\infty dt\, |A(X,Y)|^2  
\label{intensity} 
\end{eqnarray} 
In what follows we concentrate on identical final states $X=Y=\pi^+\pi^-$, 
because as we shall argue they are the most sensitive channels to probe 
the novel effects associated with the CPTV parameter $\omega$. 
Indeed~\cite{pdg}  the amplitudes 
of the 
CP violating decays $K_L \to \pi^+\pi^-$ are 
suppressed by factors of order ${\cal O}(10^{-3})$, 
as compared to the principal
decay mode of $K_S \to \pi^+\pi^-$.
In the absence of CPTV $\omega$, (\ref{bbar}), due to the 
$K_SK_L$ mixing, such decay rates would be suppressed. This would not 
be the case, however, when the CPTV $\omega$ (\ref{bbarcptv}) parameter 
is non zero, due to the existence 
of a separate $K_SK_S$ term in that case ((\ref{bph})). 
This implies that the 
relevant parameter for CPT violation 
in the intensity is $\omega/\eta_X$, which enhances 
the potentially observed effect.

\begin{figure}[tb]
\includegraphics[width=4.2cm]{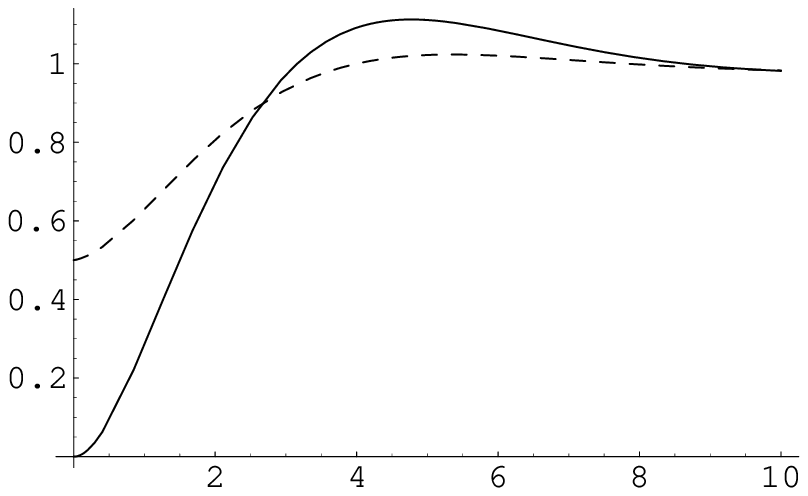}
\includegraphics[width=4.2cm]{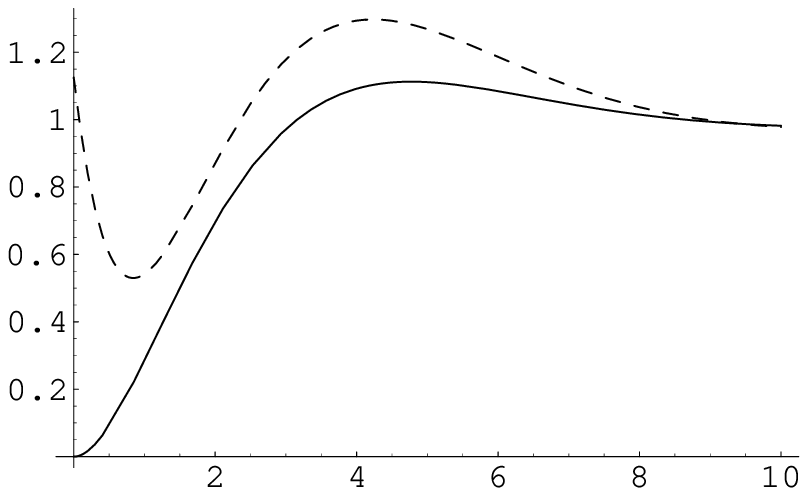}
\includegraphics[width=4.2cm]{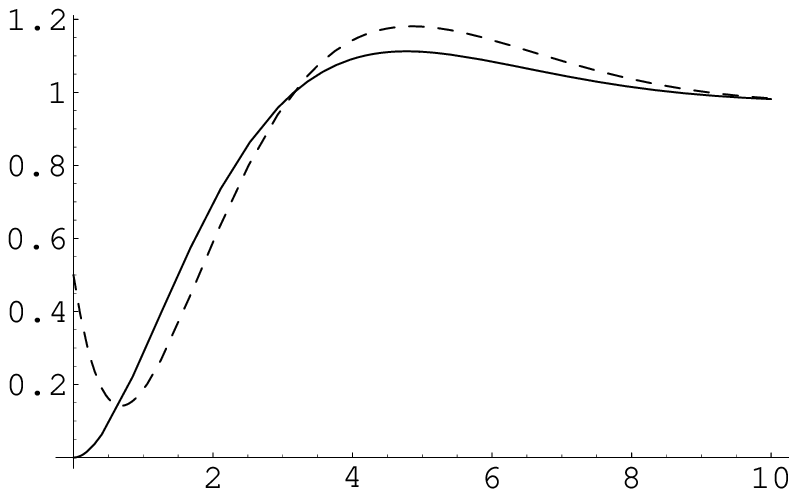}
\includegraphics[width=4.2cm]{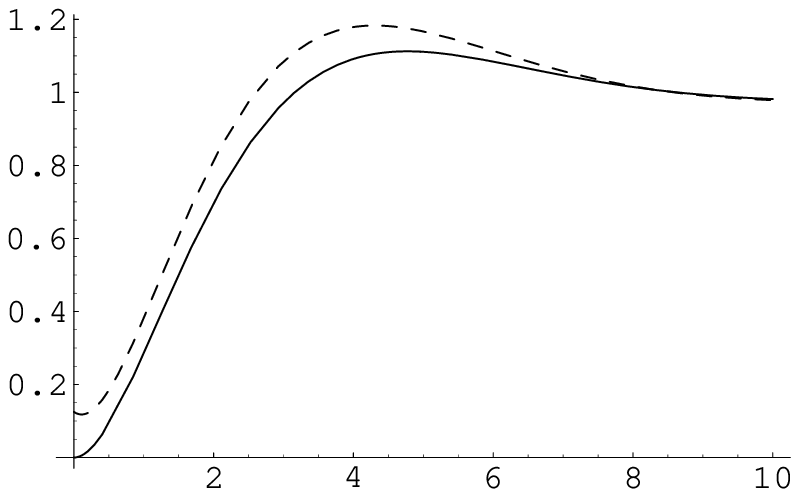}
\caption{Characteristic cases of the intensity $I(\Delta t)$, 
with $|\omega|=0$ (solid line) vs $I(\Delta t)$ 
(dashed line) with (from top left to right): (i) $|\omega|=|\eta_{+-}|$, 
$\Omega = \phi_{+-} - 0.16\pi$, (ii) $|\omega|=|\eta_{+-}|$, 
$\Omega = \phi_{+-} + 0.95\pi$, (iii) $|\omega|=0.5|\eta_{+-}|$, 
$\Omega = \phi_{+-} + 0.16\pi$, (iv) $|\omega|=1.5 |\eta_{+-}|$, 
$\Omega = \phi_{+-}$. $\Delta t$ is measured in units of $\tau_S$ (the 
mean life-time of $K_S$) and 
$I(\Delta t)$ in units of 
$|C|^2 |\eta_{+-}|^2 |\langle\pi^{+}\pi^{-}|K_S\rangle|^4 \tau_S$.} 
\label{intensityfigure} 
\end{figure}
Substituting in Eq.(\ref{intensity})
 $|A(\pi^{+}\pi^{-},\pi^{+}\pi^{-})|^2 =  
|\langle\pi^{+}\pi^{-}|K_S\rangle|^4 
( |A_1|^2  +  |A_2|^2 
 + 2 \Re e  \{A_1 A_2^{*}\} )$
and integrating over $t$ we obtain 
\be  
I(\Delta t) = |\langle\pi^{+}\pi^{-}|K_S\rangle|^4 
|C|^2 |\eta_{+-}|^2 \bigg[
I_1  + I_2  +  I_{12} \bigg]
\ee
with 
\begin{eqnarray}
I_1 (\Delta t) &=& 
\frac{e^{-\Gamma_S \Delta t} + e^{-\Gamma_L \Delta t} 
- 2 e^{-(\Gamma_S+\Gamma_L) \Delta t/2} \cos(\Delta M \Delta t)}
{\Gamma_L+\Gamma_S} 
\nonumber \\  
I_2 (\Delta t) &=& \frac{|\omega|^2 }{|\eta_{+-}|^2} 
\frac{e^{-\Gamma_S \Delta t} }{2 \Gamma_S}   
\nonumber \\  
I_{12} (\Delta t) &=& - \frac{4}{4 (\Delta M)^2 + 
(3 \Gamma_S + \Gamma_L)^2}
\frac{|\omega|}{|\eta_{+-}|}\times
\nonumber \\ 
&&\bigg[ 2 \Delta M 
\bigg( e^{-\Gamma_S \Delta t} \sin(\phi_{+-}- \Omega) -
\nonumber \\ 
&& e^{-(\Gamma_S+\Gamma_L) \Delta t/2}
\sin(\phi_{+-}- \Omega +\Delta M \Delta t)\bigg)
\nonumber \\  
&&  - (3 \Gamma_S + \Gamma_L) 
\bigg(e^{-\Gamma_S \Delta t} \cos(\phi_{+-}- \Omega) -
\nonumber \\ 
&& e^{-(\Gamma_S+\Gamma_L) \Delta t/2}
\cos(\phi_{+-}- \Omega +\Delta M \Delta t)\bigg)\bigg]
\nonumber  
\end{eqnarray}
where we have   
set $\Delta M = M_S - M_L$ and $\eta_{+-}= |\eta_{+-}| e^{i\phi_{+-}}$. 

The effects of the CPTV $\omega$ on such 
intensities $I(\Delta t)$ are indicated in figure 
\ref{intensityfigure}. The order of the CPTV effect 
is highly model dependent, and hard to evaluate. 
In line with other generic approaches to QG-decoherening
evolution~\cite{ehns,lopez,huet}, which is also 
associated with an ill definition of the concept of a $CPT$ operator,
and thus of the
antiparticle,  
in view of the lack of a well-defined scattering matrix~\cite{wald}, 
one might expect situations in which $\omega$ is of similar order 
as, say, the QG-decoherening (dimensionless) parameters~\cite{ehns,lopez}
${\widehat \alpha},{\widehat \beta}, {\widehat \gamma}$,
where  the ${\widehat \dots}$ implies division of the corresponding parameter
(with dimensions of energy) with $\Delta \Gamma = \Gamma_S - \Gamma_L$.
In optimistic scenaria of QG-induced decoherening situations, 
the relevant effects are of order $E^2/M_{QG}$, 
where $E$ a typical average energy of the Kaon system (or rest-mass, 
in the Lorentz-invariant case), 
and $M_{QG}$ the QG scale (which could be taken to be 
the Planck scale $M_P \sim 10^{19}$ GeV). For Kaons, such effects
imply that the dimensionless (hatted) quantities are expected to be of 
order $10^{-3}-10^{-4}$, thereby being well within the sensitivity of
$\phi$ factories~\cite{dafne}. Indeed,  
with $|\omega|\sim 10^{-3}-10^{-4}$ 
the new CPTV effects  become comparable to 
those associated with
$|\eta_{+-}| \sim 10^{-3}$; therefore, a precision of $10^{-3}$ 
in $I(\Delta t)$, which is needed in order 
to observe $\epsilon'$ effects, would probe sensitivities up to     
$|\omega|\sim 10^{-6}$. It is understood that a similar analysis 
can be done for the $X=Y=\pi^0\pi^0$ case.

We continue with a brief discussion concerning the distinguishability 
of the $\omega$ effect (\ref{bbarcptv}),(\ref{bph}) from 
non-quantum mechanical effects associated with the evolution, as 
in \cite{ehns}. 
The  $\omega$ effect 
can be distinguished from those 
of the QG-decoherening evolution parameters $\alpha,\beta,\gamma$, when the formalism 
is applied  to the entangled states $\phi$~\cite{huet,benatti}. 
A non-quantum mechanical evolution of the entangled Kaon state 
with $\omega =0$ has been considered in \cite{huet}.
In such a case the resulting density-matrix $\phi$ state 
$\tilde{\rho}_{\phi} ={\rm Tr}|\phi><\phi|$ can be written as
\begin{eqnarray} 
\tilde{\rho}_{\phi} &=&   
\rho_{S} \otimes \rho_{L} + 
\rho_{L} \otimes \rho_{S} 
- \rho_{I}\otimes \rho_{{\overline I}} 
- \rho_{{\overline I}}\otimes \rho_{I} 
\nonumber\\
&-& \frac{2\beta}{d} (\rho_I \otimes \rho_S + \rho_S \otimes \rho_I ) 
- \frac{2\beta}{d^*} (\rho_{{\overline I}} \otimes \rho _S + 
\rho_S \otimes \rho_{{\overline I}} )  
\nonumber\\
&+& \frac{2\beta}{d} ( \rho_{{\overline I}} \otimes \rho_L + 
\rho_L \otimes \rho_{{\overline I}}) + 
\frac{2\beta}{d^*} ( \rho_I \otimes \rho_L + \rho_L \otimes \rho_I ) 
\nonumber\\
&-&  \frac{i\alpha}{\Delta M} 
( \rho_{I}\otimes \rho_I - \rho_{{\overline I}}
\otimes  \rho_{{\overline I}})
-\frac{2\gamma}{\Delta \Gamma} 
(\rho_S \otimes \rho_S - \rho_L \otimes \rho_L)
\nonumber
\end{eqnarray}
where the standard notation $\rho_{S} = |S><S|, ~\rho_L = |L><L|,
~\rho_I = |S><L|, ~\rho_{{\overline I}} = |L><S|$ has been employed, 
$d = -\Delta M  + i \Delta \Gamma/2$,
and an overall multiplicative factor of 
$\frac{1}{2}
\frac{(1 + 2|\epsilon|^2)}
{1 - 2|\epsilon|^2{\rm cos}(2\phi_\epsilon)}$ has been suppressed.
On the other hand, the corresponding density matrix description 
of the $\phi$ state (\ref{bph}) in our case reads: 
\begin{eqnarray} 
\rho_\phi &=& 
\rho_S \otimes \rho_L + \rho_L \otimes \rho_S 
- \rho_{I}\otimes \rho_{{\overline I}} - \rho_{{\overline I}}\otimes \rho_I 
\nonumber\\
&-& \omega (\rho_I \otimes \rho_S - \rho_S \otimes \rho_I ) 
- \omega^* (\rho_{{\overline I}} \otimes \rho _S - 
\rho_S \otimes \rho_{{\overline I}} ) 
\nonumber\\
&-& 
\omega ( \rho_{{\overline I}} \otimes \rho_L - 
\rho_L \otimes \rho_{{\overline I}})  
- \omega^* ( \rho_I \otimes \rho_L - \rho_L \otimes \rho_I ) 
\nonumber\\
&-& |\omega|^2 ( \rho_{I}\otimes \rho_I + \rho_{{\overline I}}
\otimes  \rho_{{\overline I}}) + 
|\omega|^2 (\rho_S \otimes \rho_S + \rho_L \otimes \rho_L)  
\nonumber
\end{eqnarray}
with the same multiplicative factor suppressed. 
It is understood that the evolution of both $\tilde{\rho}_{\phi}$
and $\rho_\phi$ is governed by the rules given in ~\cite{ehns,lopez,huet}.
As we can see by comparing the two equations, 
the terms linear in $\omega$ in our case 
are {\it antisymmetric} under the exchange of particle states 
$1 $ and $2$, {\it in contrast} to the {\it symmetry}
of the corresponding terms linear in $\beta$ in the case of \cite{huet}.
Similar differences characterize  the terms proportional 
to $|\omega|^2$, and those proportional to $\alpha$ and $\gamma$, 
which involve $\rho_I \otimes \rho_I$, 
$\rho_{\overline I} \otimes \rho_{\overline I}$,
$\rho_S\otimes \rho_S$, $\rho_L\otimes \rho_L$. 
Such differences are therefore important 
in disentangling the $\omega$ CPTV effects proposed here 
from non-quantum mechanical evolution effects~\cite{ehns,lopez,huet,benatti}.

We next comment on the distinguishability 
of the $\omega$ effect from conventional background effects. 
Specifically, the mixing of the initial state due to 
the non-identity of the antiparticle to  
the corresponding 
particle state 
has similar form 
to that induced by a non-resonant background 
with $C=+$~\cite{dunietz}. This latter effect is known to have a small size; 
estimates based on unitarity bounds give a size  
of many orders of magnitude smaller than the 
$C=-$ effect in the $\phi$ decays ~\cite{dunietz,dafne}. 
Terms of the type $K_S K_S$ (which dominate over $K_L K_L$) coming from
the $\phi$-resonance as a result 
of CPTV 
can be distinguished from those coming from the $C=+$ background 
because they interfere differently  
with the regular $C=-$ resonant
contribution (i.e. Eq.(\ref{bph}) with $\omega=0$). Indeed, 
in the CPTV case, the $K_L K_S$ and  $\omega K_S K_S$ terms 
have the same dependence on the center-of-mass energy $s$ 
of the colliding particles producing the 
resonance,
because both terms originate from the  $\phi$-particle. Their
interference, therefore, being proportional to the real part of the 
product of the corresponding amplitudes, still displays a peak at the 
resonance. 
On the other hand, the amplitude of the $K_S K_S$ coming from the $C=+$ background 
has no appreciable dependence on $s$ and has practically vanishing imaginary part.
Therefore, given that the real part of a Breit-Wigner amplitude vanishes at the 
top of the resonance, this implies that the  
interference of the $C=+$ background  with the regular $C=-$ resonant contribution 
vanishes at the top of the resonance,
with opposite signs on both sides of the latter.    
This clearly distinguishes experimentally the two cases.

Finally we close with a comment on the 
application of this formalism to the $B$ factories.
Although, formally, the situation is identical to the one discussed above, however
the sensitivity of the CPTV $\omega$ effect for the $B$ system is much smaller.
This is due to 
the fact that 
in $B$ factories there is no particularly ``good'' 
channel $X$ (with $X=Y$) for which 
the corresponding $\eta_X$ is small. 
The analysis in that 
case may therefore be performed in
the equal sign 
dilepton channel, where the 
branching fraction is more important, and a high statistics 
is expected. Results will 
appear in future work.

{\it Acknowledgments:} This work has been supported by the 
CICYT Grant FPA2002-00612
and by the European Union (contract HPRN-CT-2000-00152).

\end{document}